\documentclass[aps,english,prd,showpacs,showkeys,twocolumn,longbibliography, nofootinbib]{revtex4-1}

\usepackage[english]{babel}
\usepackage[utf8]{inputenc}
\usepackage[T1]{fontenc}
\usepackage{fnpct} 
\usepackage{ulem} 
\usepackage{scrextend} 
\usepackage{enumitem}

\usepackage{amsmath}
\usepackage{amsfonts}
\usepackage{amssymb}
\usepackage{mathtools, stmaryrd}
\usepackage{empheq}
\usepackage{tensor}

\usepackage{fancyvrb}
\usepackage{graphicx}
\usepackage{floatrow}
\usepackage[usenames,dvipsnames]{xcolor}
\usepackage{tabularx}

\input{aastexv6.sty} 

\usepackage{url}
\usepackage[pdftex,colorlinks=true, pdfstartview=FitV, linkcolor= Mlink, citecolor=Mlink, urlcolor= Mlink, hyperindex=true, hyperfigures=true]{hyperref}
\hypersetup{linktoc=page}

\input{aastexv6.sty} 
\input{Short_cut.sty}

\begin{document}

\title[Is backreaction in cosmology a relativistic effect?]{Is backreaction in cosmology a relativistic effect? On the need for an extension of Newton's theory to non-Euclidean topologies}

\author{Quentin Vigneron$^{1,2}$}

\address{$^1$ Institute of Astronomy, Faculty of Physics, Astronomy and Informatics, Nicolaus Copernicus University, Grudziadzka 5, 87-100 Toru\'n, Poland}
\address{$^2$ Univ Lyon, Ens de Lyon, Univ Lyon1, CNRS, Centre de Recherche Astrophysique de Lyon UMR5574, F–69007, Lyon, France}
\email{quvigneron@gmail.com, quentin.vigneron@ens-lyon.fr}
\vspace{10pt}
\date{\today}

\begin{abstract}

Cosmological backreaction corresponds to the effect of inhomogeneities of structure on the global expansion of the Universe. The main question surrounding this phenomenon is whether or not it is important enough to lead to measurable effects on the scale factor evolution eventually explaining its acceleration or the Hubble tension. One of the most important result on this subject is the {\BET} (Buchert \& Ehlers, 1997) stating that backreaction is exactly zero when calculated using Newton's theory of gravitation, which may not be the case in general relativity. It is generally said that this result implies that backreaction is a purely relativistic effect. We will show that this is not necessarily the case, in the sense that this implication does not apply to a universe which is still well described by Newton's theory on small scales but has a non-Euclidean topology. The theorem should therefore be generalised to account for such a scenario. In a heuristic calculation where we construct a theory which is locally Newtonian but defined on a non-Euclidean topology, we show that backreaction is non-zero, meaning that it might be non-relativistic depending on the topological class of our Universe. However, that construction is not unique and remains to be justified from a non-relativistic limit of general relativity.

\end{abstract}

\maketitle

\section{Introduction}

The cosmological principle asserts that there exists a scale above which the Universe can be considered as homogeneous and isotropic. The Standard Model of Cosmology then assumes that the inhomogeneities under this scale do not affect the dynamics of domains of size bigger than the homogeneity scale. This implies that the expansion of such domains is given by the (homogeneous) Friedmann-Lema\^itre-Robertson-Walker (FLRW) solution of general relativity, and the Friedmann expansion laws. That is why the dynamics of structures, under the standard model, is solved as a deviation around an FLRW background expansion. However, this assumption is not a consequence of the cosmological principle and is an additional hypothesis. In reality the inhomogeneities (with scales typically smaller than the homogeneity scale) might affect the expansion at large scales. This effect is called \textit{the cosmological backreaction}. The main question surrounding this phenomenon is whether or not it is important enough to lead to measurable effects on the scale factor evolution eventually explaining its acceleration \cite{2008_Buchert} or the Hubble tension \cite{2020_Heinesen_et_al}.

One of the main and most fundamental result obtained about backreaction, known as the \textit{\BET} \cite{1997_Buchert_et_al}, states that, when calculated using Newton's theory of gravitation, the cosmological backreaction is exactly globally zero (when periodic boundary conditions are assumed), which is not necessarily the case when using general relativity. This has two consequences widely accepted today (e.g. \cite{2018_Buchert}): the study of the backreaction phenomenon has to be done using general relativity (e.g. \cite{Back_I,Back_II, 2020_BMR}); if our Universe is well described locally by the Newtonian dynamics, then the backreaction of the inhomogeneities on the expansion should be negligible and could not explain the dark energy. In other words, the backreaction is a relativistic phenomenon, meaning that in a Newtonian limit of general relativity, it should arise only at post-Newtonian orders.

The goal of this paper is to show that this last statement \textit{might} not be true as the {\BET} has two limitations, which need to be solved in order to show whether or not the backreaction is really a relativistic effect:
\begin{enumerate}
	\item the theorem relies on a description of expansion in Newton which differs from the one in general relativity,
	\item the interpretation of the theorem is only valid for a model universe whose topology lies in the Euclidean class of topologies (see section~\ref{sec:class_top} for precisions on the definition of that term), as Newton's theory is defined on such a topology.
\end{enumerate}
In particular, solving the second limitation requires an extension of Newton's theory to non-Euclidean classes of topologies such that it can be retrieved by a non-relativistic limit of general relativity, i.e. with $c\rightarrow\infty$. This paper does not intend to propose such an extension, but rather give arguments for its search, which is a new problem of cosmology and general relativity.

In the following, we will first introduce the relativistic and Newtonian approaches to describe backreaction (section~\ref{sec::BET}). We will then present the two limitations of the {\BET} in sections~\ref{sec::BET?} and \ref{sec::top}. We will see that the first one has already been solved by \cite{2021_Vigneron}, and we will propose a heuristic solution to the second one.

\section{The cosmological backreaction}
\label{sec::BET}

\subsection{Backreaction in general relativity}
\label{sec::Back_GR}

Quantifying the effects of small-scale inhomogeneities on higher scales requires an averaging procedure. This has been developed for dust irrotational fluids in \cite{Back_I}, perfect irrotational fluids in \cite{Back_II} and general rotational fluids in \cite{2020_BMR}. For the purpose of this paper we will only consider a dust irrotational fluid.

The formalism is based on averages of scalar quantities performed on a spatial domain $\CD$ comoving with the fluid in its orthogonal foliation. The averaging procedure on a scalar $\psi$ is defined as $\average{\psi}{\CD}(t) \coloneqq \frac{1}{V_\CD}\int_\CD \psi \sqrt{\mathrm{det}(h_{ab})}\dd^3 x$, where $\T h$ is the spatial metric and $V_\CD \coloneqq \int_\CD \sqrt{\mathrm{det}(h_{ab})}\dd^3 x$ the volume of $\CD$.

The goal is then to find an evolution equation for the expansion of the volume $V_\CD$ and compare it with the Friedmann expansion laws. As inhomogeneities are present inside $\CD$, the averaging procedure allows to quantify their effect at the scale of $\CD$. The domain $\CD$ is arbitrary but should be bigger than the homogeneity scale to properly quantify the backreaction on the global expansion of the Universe. However as the boundary conditions of $\CD$ are still unknown, it is generally assumed that the model universe is spatially closed\footnote{We stress that ``closed'' does not necessarily mean spherical as there exists multi-connected non-spherical topologies which are closed, including with a negative scalar curvature.} and that the averaging procedure is performed over the whole volume of that universe. This choice of boundary conditions is the most physical in a cosmological context. From now on we will consider such a model universe and we denote its spatial sections as $\Sigma$.

Then performing the averaging procedure on the scalar parts of the Einstein equation, and introducing the scale factor $a(t) \coloneqq V_\Sigma^{1/3}$ which quantifies the expansion of $\Sigma$, one finds the following expansion laws:
\begin{align}
	3\left(\frac{\dot{a}}{a}\right)^2	- 8\pi G \frac{M_\Sigma}{V_\Sigma} + \frac{\Saverage{\Rt}}{2} - \Lambda &= -\frac{\CQ_\Sigma}{2}, \label{eq::Intro_Buch_2} \\
	3\frac{\ddot{a}}{a} + 4\pi G\frac{M_\Sigma}{V_\Sigma} - \Lambda &= \CQ_\Sigma, \label{eq::Intro_Buch_1}
\end{align}
where
\begin{equation}
	\CQ_\Sigma \coloneqq \average{\theta^2}{\Sigma}- \frac{2}{3}\average{\theta}{\Sigma}^2 - \average{\tensor{\Theta}{_c_d}\tensor{\Theta}{^c^d}}{\Sigma}. \label{eq::QD_rel}
\end{equation}
$\Rt$ is the spatial scalar curvature, $M_\Sigma$ is the total mass in $\Sigma$, $\theta \coloneqq \Theta^{cd}h_{cd}$ with $\T\Theta$ the expansion tensor of the cosmic fluid.

Equations~\eqref{eq::Intro_Buch_2} and \eqref{eq::Intro_Buch_1} are similar to the Friedmann expansion laws, but with the additional effective source term $\CQ_\Sigma$. This term is related to the inhomogeneities in $\Sigma$ via the expansion tensor $\T\Theta$ and is a source for the acceleration of the scale factor. In this sense this term quantifies the effects of inhomogeneities on the expansion of $\Sigma$, and therefore is called \textit{the cosmological backreaction term}. In part because the averaging procedure is only well-defined for scalars, the vector and tensor parts of the Einstein equation are missing and the system \eqref{eq::Intro_Buch_2}-\eqref{eq::QD_rel} is not closed. In particular the time dependence of $\CQ_\Sigma$ is unknown.

\subsection{Backreaction in Newton's theory}
\label{sec::Back_Newt}

In the classical formulation of Newton's theory, expansion is not present at a fundamental level. To be able to apply this theory in a cosmological context the velocity $\T v$ is decomposed into an expansion velocity\footnote{Also called Hubble flow, or homogeneous deformation.} $\T{v}_{\text{H}}$ with ${v}_{\text{H}}^a \coloneqq H(t) \, x^a$ (in Cartesian coordinates), and a peculiar-velocity $\T P$: we have $\T v = \T{v}_{\text{H}} + \T P$, \citep[e.g.][]{1980_Peebles}. The peculiar-velocity along with the mass density are periodically defined on the absolute Euclidean space $\mE^3$. The periodicity corresponds to a cubic domain, denoted $\PerD(t)$, whose expansion rate is $H$.

Because the absolute Euclidean space is periodically tiled with the domain $\PerD$, a natural interpretation of this decomposition is to describe an effective flat 3-torus which is expanding, i.e. a closed Universe in expansion. In this interpretation, once the expansion velocity is introduced, the spatial velocity which describes the fluid is considered to be the peculiar-velocity $\T P$. Therefore, this construction allows for the description of global expansion via $\T{v}_{H}$, and inhomogeneities via $\T P$.

In this approach the expansion rate $H(t)$ is not prescribed. The goal is then to use Newton's equations to find an evolution equation for this variable. This was done by \citet{1997_Buchert_et_al}, where the authors use the same averaging formalism described in the previous section for general relativity. Then, when the averaging operator is applied to the Raychaudhuri equation, one obtains the following evolution equation:
\begin{equation}
3\frac{\ddot a}{a} + 4\pi G \frac{M_\PerD}{V_\PerD(t)} - \Lambda = \CQ_\PerD, \label{eq::Exp_law}
\end{equation}
with
\begin{equation}
	\CQ_\PerD \coloneqq \average{\theta^2}{\PerD} - \frac{2}{3}\average{\theta}{\PerD}^2 - \average{\tensor{\Theta}{_c_d}\tensor{\Theta}{^c^d}}{\PerD} + \average{\tensor{\Omega}{_c_d}\tensor{\Omega}{^c^d}}{\PerD}, \label{eq::QD}
\end{equation}
where $\tensor{\Theta}{_a_b} \coloneqq \nabla_{(a}v_{b)}$ and $\tensor{\Omega}{_a_b} \coloneqq \nabla_{[a}v_{b]}$ are respectively the expansion and the vorticity tensors of the velocity $\T v$, with $\theta \coloneqq \tensor{\Theta}{_c_d}h^{cd}$ and $\nabla_a$ the spatial covariant derivative related to the flat metric $h_{ab}$.

Similarly to the relativistic case we retrieve the Friedmann-like expansion law~\eqref{eq::Intro_Buch_1} for the acceleration of the scale factor with, apparently, an additional effective source term $\CQ_\PerD$ depending on the inhomogeneities in $\PerD$\footnote{The form of the first equation~\eqref{eq::Intro_Buch_2} can also be retrieved by time integrating equation~\eqref{eq::Exp_law}.}. But while in the relativistic case the expansion tensor $\T\Theta$ present in $\CQ_\Sigma$ is unknown without additional assumptions on the behaviour of the inhomogeneities, in the Newtonian case this tensor along with the vorticity tensor is explicitly known as function of the expansion rate $H$ and the peculiar velocity $\T P$ with: 
\begin{equation}
	\tensor{\Theta}{_a_b} =  H h_{ab} + \nabla_{(a}P_{b)} \quad ; \quad \tensor{\Omega}{_a_b} = \nabla_{[a}P_{b]}. \label{eq::theta_omega}
\end{equation}
This is a major difference as it implies that  the cosmological backreaction in Newton is \textit{a boundary term}:
\begin{equation}
	\CQ_\PerD \coloneqq \average{\nabla_c\left(P^c \nabla_dP^d - P^d\nabla_dP^c\right)}{\PerD} - \frac{2}{3}\average{\nabla_cP^c}{\PerD}^2.
\end{equation}
Because $\T P$ is by definition periodic on the domain $\PerD$, then this equation implies ${\CQ_\PerD = 0}$. This is the \textit{\BET}\footnote{It is also possible to describe an anisotropic global expansion by choosing $\T{v}_{H}^a \coloneqq H x^a + \delta^{ac}H_{cd}x^d$, with $H_{ab}(t)$ a symmetric traceless matrix (see Appendix~B of \cite{1997_Buchert_et_al}). In this case a non-zero backreaction appears, with $\CQ_{\PerD} \coloneqq - \average{H_{ cd}H^{ cd}}{\PerD}$. However, it is necessarily negative, hence it cannot mimic dark energy and as for now there is no evidence for a substantial anisotropy in the expansion of our Universe. Furthermore this ``backreaction'' does not correspond to a backreaction of inhomogeneities as it only depends on the global anisotropy of the expansion, for which it is not possible to obtain a constraint equation from Newton's equations \cite{2021_Vigneron}. \cite{1997_Buchert_et_al} also considers a more general expansion velocity with a non-symmetric tensor $H_{ab}$. This has been proven to be incompatible with general relativity in \cite{2021_Vigneron} (see Appendix~B where $\omega_{ab}$ plays the role of $H_{ab}$).}. This result is true for any inhomogeneous solution taken for $\T P$ and any type of cosmic fluid. Because ${\CQ_\PerD = 0}$ is a consequence of a boundary term, then this result is also true for any closed topology of the Euclidean class (see section~\ref{sec:class_top}), and not only for the 3-torus.

As stated in the introduction, the {\BET} seems to imply that the backreaction phenomenon is a purely relativistic effect, meaning that its study requires either post-Newtonian considerations or the full general relativity. We will see in the next sections that this statement can be questioned, first because the theorem relies on a description of expansion that differs from the one in general relativity, second because its physical implications are restricted to a model universe with a Euclidean topology.

\section{Is the {\BET} compatible with general relativity?}
\label{sec::BET?}

In the relativistic calculation (section~\ref{sec::Back_GR}) the expansion corresponds to the growth rate of the total volume of a closed universe. In the Newtonian case, the expansion arises from a construction where the velocity $\T v$ is decomposed into an expansion velocity and a peculiar velocity, the expansion corresponding to the growth rate of the volume of the periodic domain on which $\T P$ is defined. This description of expansion fundamentally differs from the relativistic case as the manifold on which Newton's equations are defined is not closed, but is the (infinite) Euclidean space $\mE^3$. Indeed, while $\T P$ could be defined on a closed space with $\mE^3$ being the covering space, the expansion velocity $\T{v}_H$ cannot because it is not periodic in $\mE^3$. So the interpretation in terms of an expanding closed space is only effective. Then because expansion is not described in the same way in the two theories, and because in the Newtonian case it arises from a construction not yet justified from general relativity, this questions the compatibility with the latter theory of the results discussed in section~\ref{sec::Back_Newt}, especially the {\BET}.

A solution would be to suppose that the space on which Newton's equations are defined is closed and derive an expansion law for the volume of this space. However, the Poisson equation would imply that the mass density should be zero everywhere, which can be seen when integrating this equation over the volume of the closed space~\cite{2020_Barrow}. Therefore from the classical formulation of Newton's theory it is not possible to have a description of expansion which would match the one of general relativity. This strengthens the question raised in the previous paragraph and therefore the need for a derivation of the {\BET} as a limit of general relativity.

We performed such a derivation in \cite{2021_Vigneron} using the Newton-Cartan formalism. This formalism is the closest formulation of Newton's theory to general relativity, both in terms of equations and conceptually (see \cite{1972_Kunzle} for a complete formulation of the Newton-Cartan theory). Furthermore, the Newton-Cartan equations have been derived from general relativity using a well-defined non-relativistic covariant limit \cite{1976_Kunzle}. This contrasts with the usual way of performing the Newtonian limit which is generally done using specific coordinate systems (see e.g. \cite{1995_Kofman_et_al}). Hence, we can consider that retrieving the conclusion of the {\BET} from the Newton-Cartan formulation ensures the compatibility of that result with a non-relativistic limit of general relativity.

In \cite{2021_Vigneron}, we performed a 1+3 decomposition of the Newton-Cartan equations to obtain covariant 3-dimensional equations, \textit{the 1+3-Newton-Cartan equations}. From this system of equations an expansion law is derived (see equation~(52) in \cite{2021_Vigneron}). This law is equivalent to the Friedmann equation if the isotropy of the expansion is assumed, i.e. $\T\Xi = 0$ in \cite{2021_Vigneron}. In particular it features no backreaction, thus retrieving the conclusion of the {\BET}. A fundamental aspect in this derivation is the fact that the expansion and vorticity tensors of the cosmic fluid acquire the same form as in equation~\eqref{eq::theta_omega}, but with a major difference: the term $H h_{ab}$ does not arise from the gradient of a vector as in the classical case, but from the unique scalar-vector-tensor decomposition of $\T\Theta$ (see \cite{2008_Straumann} for the proof of the decomposition theorem), hence $H(t)$ is a fundamental field of the theory. Furthermore, we showed in \cite{2021_Vigneron} that this term represents the expansion rate of the volume of a closed space, hence reconciling the Newtonian description of expansion with the relativistic one.

\section{The issue of topology}
\label{sec::top}

While the compatibility of the {\BET} with general relativity seems to imply that cosmological backreaction is a relativistic effect, we will see in this section that taking into account the topological class of the Universe {questions} this statement.

\subsection{Classes of topology}
\label{sec:class_top}

We precise in this section some notions of topology that are used in the remaining of this paper.

In 3-dimensions, the classification of the closed 3-manifolds is given by Thurston's classification \citep{1982_Thurston, 2006_Morgan_et_al}, which says that the topology of any closed differentiable 3-manifold can be decomposed into a connected sum of pieces that each has one of the eight Thurston topologies. Among those eight classes of topology we have in particular the Euclidean, spherical and hyperbolic topologies. A topology is said to be Euclidean (respectively spherical; hyperbolic) if its universal cover is homeomorphic to $\mE^3$ (respectively to $\mS^3$; to $\mH^3$) and whose fundamental group is a discrete subgroup of $\mathbb{R}^3 \times SO(3)$ (respectively of $SO(4)$; of $PSL(2,\mathbb{C})$). These three classes of topology are the only one where an isotropic Ricci curvature tensor can be defined, which is the reason why they are the only one used in the $\Lambda$CDM model. For the description of the other five classes of topologies, see \citet{1995_La_Lu}.
Then the term `Euclidean topology' we use throughout the present paper refers to the topology of a 3-manifold that lies in the Euclidean class of Thurston's decomposition, while a `non-Euclidean topology' refers to all the other possible 3-dimensional topologies.

This notion of `Euclidean' and `non-Euclidean' is not to be confused with Euclidean and non-Euclidean geometries which generally refer, in the literature about cosmology and general relativity, to the presence or not of a non-zero Ricci curvature tensor.
In particular, while Euclidean geometry implies by definition a zero Ricci tensor, a 3-manifold whose topology lies in the Euclidean class can have a non-zero Ricci tensor. This is not the case in Newton's theory, where the topology is Euclidean and the curvature is zero, but this is the case for some solutions of General relativity \citep[see for instance the relativistic numerical simulations of][which are performed on a 3-torus, i.e. a Euclidean topology, but where the spatial Ricci tensor is non-zero]{2019_MacPherson_et_al}. The same applies for spherical and hyperbolic topologies, for which the Ricci tensor is not necessarily of the form $\Rt_{ab} = {\Rt}/{3} \, h_{ab}$. In other words, the terms `Euclidean' and `non-Euclidean' used in the present paper characterise the topology of the 3-manifold and not its Ricci curvature.

Of course there are contraints on the relation between the topology and the curvature. For instance, a 3-manifold with an everywhere-zero Ricci tensor has necessarily a Euclidean topology, which is why Newton's theory is defined only for Euclidean topologies. This implies that it is not possible to define an everywhere-zero Ricci tensor on a non-Euclidean topology, hence the need for a non-zero Ricci tensor in the spherical or hyperbolic cases.

\subsection{The need for a non-Euclidean Newtonian theory}
\label{sec:Need_NEN}

To apply the conclusions of the {\BET} to our Universe, the latter would have to be well described by Newton's theory on small scales, but also on the global scale by having the same kind of topology as the one of that theory, i.e. a Euclidean topology, along with the closedness of space. However, general relativity \textit{a priori} allows for any of the closed 3-manifolds described by Thurston's classification. If we consider this theory to be the genuine theory of gravitation, then our Universe could have a non-Euclidean topology, but still being well-described locally by Newton's theory. In that case, the {\BET} could not be used to estimate the smallness or not of backreaction. Therefore, we need a generalisation of this theorem to non-Euclidean topologies, i.e. valid for a model universe which is locally Newtonian but with a non-Euclidean topology. Such a generalisation requires the construction of an extension of Newton's theory to non-Euclidean topologies. We call such an extension: a \textit{non-Euclidean Newtonian theory}. By definition, this theory should be equivalent to Newton's theory on scales small with respect to the (finite) volume of the model universe. In addition to generalising the Buchert-Ehlers theorem, such a theory would be a powerful tool to study the effects of topology on structure formation. In particular, we expect that theory to provide a theoretical framework for N-body calculation on a non-Euclidean topology.

To understand more deeply the importance of the issue of topology, we can define the following non-Euclidean Newtonian theory, using the procedure developed in \cite{2009_Roukema_et_al, 2020_Barrow}, and derive its backreaction:
\begin{enumerate}
	\item We consider a spherical or hyperbolic topology. This implies that the spatial Ricci curvature must be non-zero as stated at the end of section~\ref{sec:class_top}. For simplicity, we take it to be purely scalar with $\Rt_{ab} = {\Rt(t)}/{3} \, h_{ab}$. This choice of Ricci tensor implies a one to one relation between the topological class considered, and the sign of the scalar curvature: $\CR>0$ for a spherical topology, and $\CR<0$ for a hyperbolic topology.
	\item We algebraically keep the classical Newtonian equations in there kinematical form, especially the Raychaudhuri equation and formulas~\eqref{eq::theta_omega} for the kinematical tensors, and we assume that the spatial metric involved in the connection $\T\nabla$ is non-flat with the above curvature.
\end{enumerate} 
This means that we have this system of equations:
\begin{align}
&\Theta_{ij} = H h_{ij} + \nabla_{(i} P_{j)} \quad : \quad \Omega_{ij} = \,\nabla_{[i}P_{j]} \label{eq:kin_1}\\
 &\left(\partial_t + \mathcal{L}_{\boldsymbol{P}}\right)\rho + \rho \theta = 0 \\
&\left(\partial_t + \mathcal{L}_{\boldsymbol{P}}\right)\theta = -4\pi G\rho + \Lambda - \Theta_{ij}\Theta^{ij} + \Omega_{ij}\Omega^{ij} \,+ \nabla_i a^i_{\not= \mathrm{grav}} \\
&\left(\partial_t + \mathcal{L}_{\boldsymbol{P}}\right)\Omega_{ij} = \,\nabla_{[i} (a_{\not= \mathrm{grav}})_{j]}\\
&\Rt_{ab} = {\Rt(t)}/{3} \, h_{ab} \textrm{\ \ in\ \ } \nabla_i \label{eq:kin_2}
\end{align}
where $\T{a}_{\not= \mathrm{grav}}$ is the non-gravitational spatial acceleration underwent by the cosmic fluid and $\mathcal{L}_{\boldsymbol{P}}$ is the Lie derivative with respect to the peculiar velocity.

The theory described by these equations is therefore defined on a manifold with a non-Euclidean topology because $\Rt_{ab} \not= 0$, but reduces to Newton's theory on scales small with respect to the curvature scale, i.e. small with respect to the size of that manifold\footnote{With $\Rt_{ab} = {\Rt(t)}/{3} \, h_{ab}$, the volume of the 3-manifold is proportional to $|\CR|^{-3/2}$.}. Hence this defines a non-Euclidean Newtonian theory. We stress that the addition of a spatial curvature is not done in order to take into account relativistic effects, as in \cite{2014_Abramowicz_et_al}, but only to allow for non-Euclidean topologies, i.e. it is not a relativistic correction. Therefore we expect that if this theory can be retrieved via a non-relativistic limit of general relativity, this curvature should be the zeroth order of the real spatial curvature, contrary to the purely Newtonian case where that zeroth order vanishes. We also expect that it should play a minor role on small scales, as it is spatially constant, but as we will see might play a major role on the global scale.

In that theory, the cosmological backreaction is still given by formula~\eqref{eq::QD} (which is obtained by averaging the Raychaudhuri equation) but due to the presence of a non-zero Ricci tensor, it is not anymore zero but acquires a dependence on the peculiar velocity and the scalar curvature with
\begin{equation}
	\average{\CQ}{\PerD} = \frac{\CR}{3}\average{P^cP_c}{\PerD} \not= 0. \label{eq::QD_NEN}
\end{equation}
This non-zero term arises from a commutation of the spatial connection when inserting equation~\eqref{eq::theta_omega} into equation~\eqref{eq::QD}.

We see that the backreaction depends on the topological class via the scalar curvature~$2{\CR}/{3}$ (but not explicitly on the specific topology within a class), and also on the inhomogeneities in the model universe, via the mean specific kinetic energy $\average{P^cP_c}{\PerD}/2$. As this energy is linked to the virialisation of the structures, this second dependence is an expected behaviour of backreaction (see \cite{2008_Buchert}). For a hyperbolic model universe $\Rt <0$ and the backreaction is necessarily negative; for a spherical model universe $\Rt > 0$ and the backreaction is necessarily positive. The constant sign property of $\CQ_\PerD$ suggests that build-up effects over the life of the Universe could lead to a departure of its expansion with respect to the $\Lambda$CDM model, due to a non-Euclidean topology. However, the second order dependance of this effect on the peculiar velocity could prevent this departure to be significant enough to be measured. For instance, assuming $\Lambda$CDM cosmology and using the current upper bound for the late-Universe curvature parameter $|\Omega^0_\CR| \coloneqq \left|c^2\CR/(6H^2_0)\right| \lesssim 10^{-3}$ (from Planck's data and BAO measurements \cite{2020_Planck_VI})  and the typical estimate of the peculiar velocity with $\T P\cdot\T P/c^2 \sim 10^{-6}$, then $\left|\Omega^0_{\CQ}\right| \coloneqq \left|\CQ_\PerD/(6H_0^2)\right| \lesssim 10^{-9}$, which is negligible with respect to the contribution of matter density and cosmological constant.

Still the result~\eqref{eq::QD_NEN} means that in the non-Euclidean Newtonian theory defined above, the conclusion of the {\BET} does not hold anymore and that cosmological backreaction \textit{might} be non-relativistic depending on the topology of our Universe. 
This result is still heuristic as the non-Euclidean Newtonian theory we defined above lacks a clear justification from general relativity which should come in the form of a non-relativistic limit. Therefore we strongly emphasise that we should not take formula~\eqref{eq::QD_NEN} and its interpretation as physical before this justification arrives. Nevertheless this calculation motivates the need for the construction of a non-Euclidean Newtonian theory compatible with general relativity.

\subsection{Caveats of the procedure}
\label{sec:caveats}

The main problem with the theory defined by the system~\eqref{eq:kin_1}-\eqref{eq:kin_2} is that N-body calculation is not anymore possible. If one reintroduces the peculiar gravitational potential $\nabla^i \Phi \coloneqq -\left(\partial_t +P^j\nabla_j\right)P^i - 2HP^i$, as defined for Newtonian cosmology \citep[see][]{1980_Peebles}, in that system, then we obtain
\begin{equation}
	\Delta \Phi = 4\pi G \left(\rho - \average{\rho}{\PerD}\right) - \frac{\mathcal{R}}{3}\left(P^cP_c - \average{P^cP_c}{\PerD}\right).
\end{equation}
Similarly to the weak field limit of general relativity around a spatially curved FLRW model, this equation features an additional term, with respect to the Poisson equation, proportional to the curvature. However because that term is non-linear (contrary to the weak field limit where it is $\Phi\CR/2$), N-body calculation is not possible. In other words, the theory defined by the system~\eqref{eq:kin_1}-\eqref{eq:kin_2} looses one of the main strength Newton's theory has over general relativity.

The reason why the Poisson equation is not retrieved is because once we introduce a non-zero Ricci curvature tensor in the spatial connection, the kinematical system of Newton's theory is not anymore equivalent to the gravitational system, which is
\begin{align}
&\boldsymbol{g} = \dot{\boldsymbol{v}} - \boldsymbol{a}_{\not=\mathrm{grav}} \quad &; \quad \dot\rho + \rho \nabla_i v^i = 0 \label{eq:grav_1}\\
&\nabla_{[i}g_{j]} = 0 \quad &; \quad \nabla_ig^i = -4\pi G\rho \\
&\Rt_{ab} = {\Rt(t)}/{3} \, h_{ab} \textrm{\ \ in\ \ } \nabla_i \, . \label{eq:grav_2}
\end{align}
Hence the system~\eqref{eq:kin_1}-\eqref{eq:kin_2} and the system~\eqref{eq:grav_1}-\eqref{eq:grav_2} represent two different theories that can be said to be locally Newtonian (they reduce to Newton's theory an scales small with respect of the curvature scale) but defined on a 3-manifold with a non-Euclidean topology. This shows that the procedure of \citep{2009_Roukema_et_al, 2020_Barrow}, namely adding a non-zero Ricci tensor in Newton's equations suffers from a major caveat: it is not unique.

One might want to use the system~\eqref{eq:grav_1}-\eqref{eq:grav_2} instead of the system~\eqref{eq:kin_1}-\eqref{eq:kin_2}, as the Poisson equation is by definition present. However, in that case, it is not anymore possible to describe expansion, as defining the Hubble flow vector field ${v}_{\text{H}}^a \coloneqq H(t) \, x^a$ on a non-Euclidean topology is not possible.

\subsection{Proposed strategy}

In section~\ref{sec:Need_NEN}, we considered a theory to be Newtonian if it features algebraically the same equations, but with the presence of a non-zero spatial Ricci tensor. This ensured that on scales small with respect to the closed topological space (i.e. small with respect to $1/\sqrt{|\CR|}$), the equations reduce to Newton's equations. However this approach lead to, at least, two different non-Euclidean Newtonian theories which have both major caveats.

Following section~\ref{sec::BET?} and \citet{2021_Vigneron}, we saw that Newton's theory is best formulated in its geometrised form with Galilean manifolds, i.e. with the Newton-Cartan formulation. Therefore if one still wants to use the procedure of \citep{2009_Roukema_et_al, 2020_Barrow} to extend Newton's theory on non-Euclidean topologies, the addition of the spatial Ricci tensor needs to be done in the (spacetime) Newton-Cartan equations. In this approach a theory is said to be \textit{Newtonian} if it is defined on a Galilean manifold whatever the topological class considered. This means that the Galilean invariance is taken as a fundamental principle of the theory. This strategy is used in a follow-up paper \citep{2021_Vigneron_c}.

Ultimately, the right non-Euclidean Newtonian theory needs to be derived from general relativity in a limit where $c \rightarrow \infty$. While such a limit has been defined in various ways in the literature, often using specific coordinate systems which imposed a Euclidean topology, the use of Galilean structures proposed above suggests that the non-relativistic limit which needs to be considered is the Galilean limit of Lorentzian spacetimes \citep[e.g.][]{1976_Kunzle, 2020_Hansen_et_al}. In this limit, a family a Lorentzian structures on a 4-manifold becomes, when $c \rightarrow \infty$, a Galilean structure. Because such a structure can be defined on 4-manifolds with any spatial topology, it would be possible to obtain a non-Euclidean Newtonian theory from general relativity in this way \citep[see][]{2021_Vigneron_c}.

\section{Conclusion}

In this paper we analysed the {\BET} and showed that it has two limitations, implying that cosmological backreaction might be non-relativistic.

First, the description of expansion made in this theorem, using Newton's theory of gravitation, is different from the one of general relativity, implying an eventual incompatibility with the latter theory. We showed that this problem was already solved by \cite{2021_Vigneron}, where the {\BET} was retrieved using the Newton-Cartan formalism. As this formulation of Newton's theory is close to general relativity and has been derived from it with a covariant limit \cite{1976_Kunzle}, this ensures the compatibility of the {\BET} with the latter theory. 

Second, we noted that the interpretation of the theorem is limited to Euclidean topologies, i.e. it cannot be applied to a model universe which is well described by Newton's theory on small scales but has a non-Euclidean topology. To strengthen the importance of this statement we proposed a heuristic calculation of a ``Newtonian backreaction'' in a model universe with a non-Euclidean topology. This lead to a non-zero backreaction of the inhomogeneities on the expansion, suggesting that this phenomenon might be non-relativistic depending on the topological class of our Universe. To justify this result, one needs to develop an extension of Newton's theory to non-Euclidean topologies which can be retrieved with a non-relativistic limit of general relativity. Searching for such a theory is a new problem of general relativity and cosmology.

Follow-up work is dedicated, first to propose non-Euclidean extensions of Newton's theory directly based on the Newton-Cartan formalism and study their cosmological backreaction \cite{2021_Vigneron_c}; then to derive one of them as a limit of general relativity using the Galilean limit of general relativity.

\section*{Acknowledgements}
This work has received funding from the European Research Council (ERC) under the European Union’s Horizon 2020 research and innovation programme (Grant agreement ERC advanced Grant 740021–ARTHUS, PI: Thomas Buchert). I would like to thank Pierre Mourier for a lot of valuable comments on the manuscript and for that discussion in June 2019 which made me realise the importance of developing a non-Euclidean Newtonian theory. I also thank L\'eo Brunswic, Etienne Jaupart and Thomas Buchert for discussions and comments on the manuscript. I am thankful to the anonymous referees for their thorough evaluations.

\section*{References}
 {When available this bibliography style features three different links associated with three different colors: links to the journal/editor website or to a numerical version of the paper are in \textcolor{LinkJournal}{red}, links to the ADS website are in \textcolor{LinkADS}{blue} and links to the arXiv website are in \textcolor{LinkArXiv}{green}.}\\

\bibliographystyle{QV_mnras}
\bibliography{bib_General}

\end{document}